\documentclass[preprint,aps,showpacs,nofootinbib,eqsecnum,floatfix]{revtex4}
\usepackage{amsmath}
\usepackage{amsfonts}
\usepackage{amssymb}
\usepackage{mathrsfs}
\usepackage{color}
\usepackage{graphicx}
\usepackage{ulem}

\def\bc{\begin{center}}
\def\nno{\nonumber}
\def\ec{\end{center}}
\def\be{\begin{eqnarray}}
\def\ee{\end{eqnarray}}

\newcommand{\omits}[1]{}

\definecolor{dyellow}{rgb}{1.,0.8,.0}
\definecolor{myblue}{rgb}{.1,.1,.7}
\definecolor{dcyan}{rgb}{.0,.6,.6}
\definecolor{dmagenta}{rgb}{0.6,0.0,0.6}
\definecolor{brown}{rgb}{0.6,0.2,0.}
\definecolor{darkblue}{rgb}{.0,.0,0.5}
\definecolor{darkred}{rgb}{0.75,0.0,0.0}
\definecolor{orange}{rgb}{1.,.6,.0}
\definecolor{dorange}{rgb}{0.8,.4,.0}
\definecolor{darkgreen}{rgb}{0.0,0.6,0.0}
\definecolor{purple}{rgb}{.4,.0,.4}
\def\N{N \hspace{-0.7em}_{_{\sim}} ~}

\def\dl{\delta}
\def\eps{\epsilon}
\def\ka{\kappa}
\def\la{\lambda}

\def\si{\sigma}

\def\d#1#2{\frac{\displaystyle #1}{\displaystyle #2}}
\def\r{\partial}

\newcommand{\vect}[1]{\mbox{\boldmath $#1$}}

\newcommand\btd{\raise 2pt
\hbox{$\hat\bigtriangledown$}\hskip 1.5pt}
\newcommand\bt{\raise 2pt
\hbox{$\bigtriangledown$}\hskip 1.5pt}

\newcommand{\SR}{$SR$}
\newcommand{\GR}{$GR$}

\newcommand{\dS}{$d{S}$}
\newcommand{\M}{${\cal M}$}
\newcommand{\AdS}{${A}d{S}$}

\newcommand{\PoR}{${P}o{R}$}%principle of relativity}
\newcommand{\PoRcl}{${P}o{R}_{c l}$}%postulate on invariants $c$ and $R$}

\newcommand{\PGL}%{${P}G{L}$}%
{projective geometry}
\newcommand{\LFT}{$LFT$}
\newcommand{\R}{$\cal R$}

\newcommand{\Mink}{$Mink$}

\def\P{{\bf P}}
\def\K{{\bf K}}
\def\H{{H}}
\def\J{{\bf J}}
\def\N{{\bf N}}
\def\M{{\it M}}
\def\R{{\it R}}

\def\PRD{{\it Phys. Rev.}~{\bf D}}
\def\PRL{{\it Phys. Rev. Lett }}
\def\PLA{{\it Phys. Lett.}~{\bf A}}
\def\PLB{{\it Phys. Lett.}~{\bf B}}

\def\CQG{{\it Class. Quantum Gravity }}

\begin{document}

\title{Principle of Relativity, Dual  Poincar\'e Group and Relativistic Quadruple}

\author{{Han-Ying Guo}$^{1}$
\email{hyguo@itp.ac.cn}}

\author{{Hong-Tu Wu}$^{2}$
\email{000@amss.ac.cn}}

\affiliation{%
${}^1$ Institute of Theoretical Physics and Kavli Institute of
Theoretical Physics of China, Chinese Academy of Sciences,
   Beijing 100190, China,}

\affiliation{%
${}^2$ Institute of Mathematics, Academy of Mathematics  and
System Sciences, Chinese Academy of Sciences,
   Beijing 100190, China.}

\date{February, 2010}

\begin{abstract}
Based on the principle of relativity with two universal constants
$(c, l)$, there is  the inertial motion group $IM(1,3)\sim
PGL(5,R)$. With Lorentz group $SO(1,3)$ for isotropy, in addition to
Poincar\'e group ${\cal P}$ of Einstein's special relativity the
dual  Poincar\'e group ${\cal P}_2$ preserves the origin lightcone
$C_O$
 and
its space/time-like region $R_\pm$ appeared at common origin of
intersected Minkowski/de Sitter/anti-de Sitter space $M/D_\pm$. The
${\cal P}_2$ kinematics is on
 a pair of degenerate Einstein manifolds $M_\pm$
with $\Lambda_\pm=\pm3l^{-2}$ for $R_\pm$, respectively. Thus, there
is a Poincar\'e double $[{\cal P},{\cal
  P}_2]_{M/M_\pm}$. There is also the de Sitter
  double $[{\cal D}_+,{\cal
 D}_-]_{D_\pm}$ for \dS/\AdS\ special relativity. Further,
 related to $M/M_\pm/D_\pm$,
 there  exist  other four 
 doubles $[{\cal D}_+,{\cal P}]_{D_+/M}$, $[{\cal D}_-,{\cal P}]_{D_-/M}$,
 $[{\cal D}_+,{\cal P}_2]_{D_+/M_\pm}$, and $[{\cal D}_-,{\cal P}_2]_{D_-/M_\pm}$. These doubles
 form a relativistic quadruple $[{\cal P}, {\cal P}_2, {\cal D}_+,{\cal D}_-]_{M/M_\pm/D_\pm}$
 for three kinds of special
 relativity on $M/D_\pm$,
respectively. The \dS\ special relativity associated with the
double $[{\cal D}_+,{\cal P}_2]_{D_+/M_+}$
 should provide the consistent kinematics for cosmic scale physics of
$\Lambda_+>0$.
\end{abstract}

\pacs{%
03.30.+p, %special relativity
11.30.Cp  %Lorentz and Poincar¨¦ invariance
%02.20.Sv  %Lie algebras of Lie groups
02.40.Dr, %Euclidean and projective geometry
%04.20.Cv, %Fundamental problems and general formalism
%04.90.+e, %Other topics in general relativity and gravitation (restricted to new topics in section 04 \footnote{Section 04 is General relativity and Gravitation)
%04.50+h, %Gravity in more than four dimensions, Kaluza-Klein theory, unified field theories; alternative theories of gravity
98.80.Jk, %Mathematical and relativistic aspects of cosmology
}

\maketitle

\tableofcontents

%%%%%%%%%%%%%%%%%%%%%%%%%%%%%%%%%%%%%%%%%%%%%%%%%%%%%%%%%%%%%%%%%%%%%%%%
%%%%%%%%%%%%%%%%%%%%%%%%       Section 1        %%%%%%%%%%%%%%%%%%%%%%%%
%%%%%%%%%%%%%%%%%%%%%%%%%%%%%%%%%%%%%%%%%%%%%%%%%%%%%%%%%%%%%%%%%%%%%%%%
\section{Introduction}

Precise cosmology\cite{90s,WMAP} shows that our universe is
accelerated expanding and is not asymptotic to the Minkowski
(\Mink) space, rather quite possibly  a de Sitter (\dS) space with
a tiny positive cosmological constant $\Lambda_+>0$.  This opens
up the era of the cosmic scale physics characterized by the
$\Lambda_+$ and greatly challenges Einstein's theory of
 relativity\cite{1905, 1922} as
its solid foundation. %
 As   ``principle theory",   Einstein's principles
  must be
reexamined from the very beginning  to explore if there is new
kinematics for the cosmic scale physics. Actually, this is the case.

In order to introduce  the cosmological constant  at the principle
level, a universal invariant constant $l$ of
 length dimension should be introduced in addition to the speed of
 light $c$.  This is also the case from some general and
 simple dimensional
analysis: in space-time physics space and time coordinates should
have right invariant dimension, i.e.  $[x^i]=L, (i=1,2,3)$ and
$[t]=T$. In order to characterize them and to   make coordinate $t$
has the same dimension with $x^i$ of dimension length, a universal
constant $c$ of velocity dimension $[c]=L/T$ and another universal
constant $l$ with $[l]=L$ should be introduced, which should also be
invariant under transformations among coordinate systems. And their
realistic values must be determined by %
experiments and observations with respect to  inertial observers.
In fact, $c$ is determined as the speed of light and $l$ is hidden
in Einstein's special relativity (\SR) and  as long as the
Euclidean assumption on time and space is relaxed, the principle
of relativity (\PoR) and the postulate for the light propagation
in \SR\ \cite{1905} can be extended and weakened to the \PoR\ of
two universal constants $(c,l)$, denoted as \PoRcl.

Based on the \PoRcl, it has been found that for inertial motion
described by Newton's first law there are linear fractional
transformations with
  common denominators (\LFT s) of the inertial motion
group $IM(1,3)$ of twenty-four generators homomorphic to the 4d
real projective group $PGL(5,R)$ with algebra $\frak{im}(1,3)\cong
\frak{pgl}(5,R)$ kinematically,  and it contains homogeneous
Lorentz group ${\cal L}:=SO(1,3)$ as subgroup for relativistic
cases\cite{SRT,SRT2}. In fact, in addition to the Poincar\'e group
${\cal P}$
 with Euclid time and space translations $(H, \P_i) \in \frak{p}$ on
 \Mink-space $M$,
there is also the dual  %
 Poincar\'e group  ${\cal P}_2$ of \LFT s\footnote{It is also called the second
 Poincar\'e group in \cite{SRT, SRT2}.}
with time and space  pseudo-translations $(H',\P'_i) \in
\frak{p}_2\cong \frak{iso}(1,3)$  proportional to $l^{-2}$ that
preserves the \Mink\ origin lightcone $C_O$
 and its space/time-like region $R_\pm$.  So, related to $M$, there is
 an algebraic doublet $(\frak{p},\frak{p}_2)$. And in terms of the  linear
combinations among translations
\be\label{eq:combinatory}%
H\pm H'=\H^\pm, \quad \P_i\pm \P'_i=\P_i^\pm %
\ee%T
 with common $\frak{so}(1,3)$ isotropy, the
 Poincar\'e doublet
 leads to the
 \dS/\AdS\
doublet $(\frak{d}_+,\frak{d}_-)$ with Beltrami translations
$(\H^\pm, \P_i^\pm)$ in the Beltrami-\dS/\AdS-space $D_\pm$ of
radius $l$, respectively, and vice versa in
$\frak{im}(1,3)\cong\frak{pgl}(5,R)$\cite{SRT, SRT2}. This is very
simple and different from the
 contraction \cite{IW,Bacry} and deformation\cite{deform} approaches.
 Thus, based on the \PoRcl\ and  within $IM(1,3)\sim PGL(5,R)$
 there are three kinds of
Poincar\'e/\dS/\AdS\ invariant \SR,
respectively\cite{Lu,LZG,1950s,BdS,Duality}. With common Lorentz
isotropy and relations (\ref{eq:combinatory}), they
{automatically} form an \SR\ triple of four related groups ${\cal
P},{\cal P}_2$, ${\cal D}_+:=SO(1,4)$ and ${\cal
D}_-:=SO(2,3)$ %
with an algebraic quadruplet $\frak q:=(\frak{p},\frak{p}_2,
\frak{d}_+,\frak{d}_-)$.  It is important that these relativistic
kinematics and their combinational structures automatically appear
in  the algebra $\frak{im}(1,3)\cong \frak{pgl}(5,R)$.

In fact, by means of projective geometry method, it can be shown
that for those relativistic kinematics and their algebraic
relations there are also  their
 group and geometry counterparts with common Lorentz isotropy, i.e. ${\cal L}={\cal P}\cap{\cal
 P}_2\cap
{\cal D}_+ \cap{\cal D}_-$, with respect to  the
 \LFT s of the group $IM(1,3)\sim PGL(5,R)$.

 This can
 be seen  from the other angle, i.e. these issues can also easily be reached by
their construction  starting from the \Mink-lightcone at the
origin. Since there is the common Lorentz isotropy, as long as the
flatness of \Mink-space is relaxed, the region defined by
$\eta_{\mu\nu}x^\mu x^\nu\mp l^2\lessgtr0$, i.e. shifting the
origin lightcone equation by $\mp l^2$, leads to the domain of the
Beltrami-\dS/\AdS-space $D_\pm$ with $x^\mu$  as the Beltrami
coordinates without antipodal identification, respectively. Since
the origin lightcone in $D_\pm$ is still Minkowskian, there are
the \SR\ triple for three
kinds of \SR\  %  as a 3-simplex
 on intersected non-degenerate
\Mink/\dS/\AdS-spaces $M/D_\pm$  and four related groups with the
quadruplet $\frak q$ \cite{SRT, SRT2}. As the ${\cal
P}_2$-kinematics, there is a pair of ${\cal P}_2$
  invariant degenerate
Einstein manifolds{\footnote{Einstein manifold
means that it satisfies vacuum Einstein field equation only.}} $M_\pm:=(M^{{\cal P}_2}_\pm, {\vect g}_\pm,
{\vect g}^{-1}_\pm, { \nabla_{\Gamma_\pm}})$ with cosmological
constant $\Lambda_\pm=\pm 3 /l^2$, where
the degenerate metric ${\vect g}_\pm$ is induced from the $C_O$ as
an absolute for its space/time-like region $R_\pm$ on \Mink-space
$M$, respectively, with its formal inverse ${\vect g}^{-1}_\pm$
and the Christoffel symbol ${\Gamma_\pm}$. Then for four groups
${\cal P}/{\cal
 P}_2/
{\cal D}_\pm$ as subgroup of $IM(1,3)$, {corresponding to their
algebraic relations (\ref{eq:combinatory})} they should share the
Lorentz group ${\cal L}$
 for isotropy, and on these spaces $M/M_\pm/D_\pm$
there is the same origin lightcone structure $[C_O]$ {in the sense
of $R_\pm=M\cap M_\pm\cap D_+\cap D_-$}. So, the ${\cal
P}_2$-kinematics for $R_\pm$ induced from $C_O=\partial R_\pm$ is
also common and with common Lorentz isotropy there are six types
of
doubles, i.e. %Poincar\'e, \dS/\AdS-$\cal P$, \dS/\AdS-${\cal P}_2$ and \dS-\AdS\
$[{\cal P},{\cal
  P}_2]_{M/M_\pm}$,  $[{\cal D}_+,{\cal P}]_{D_+/M}$, $[{\cal D}_-,{\cal P}]_{D_-/M}$,
 $[{\cal D}_+,{\cal P}_2]_{D_+/M_\pm}$, $[{\cal D}_-,{\cal P}_2]_{D_-/M_\pm}$, and $[{\cal D}_+,{\cal
 D}_-]_{D_\pm}$. Thus,
based on the \PoRcl, they automatically form a relativistic
quadruple ${\frak Q}_{PoR}:=[{\cal P}, {\cal P}_2, {\cal
D}_+,{\cal D}_-]_{M/M_\pm/D_\pm}$. % with quadruplet $\frak q$.

Since there is the dual  Poincar\'e group ${\cal P}_2$ and its
kinematics for $R_\pm$ at origin, related to \Mink-space $M$, the
symmetry and geometry are dramatically changed by
the Poincar\'e double $[{\cal P},{\cal P}_2]_{M/M_\pm}$. %
In order to restore Einstein's \SR,  $l = \infty$ has to be taken.
But, for the relativistic physics as a description of reality on
free space without gravity in the universe, the %role played by the
cosmological constant $\Lambda_+$ cannot be missed.
Then the $(H', \P'_i)$ can be  ignored up to %the lowest bound from the $\Lambda$
 \be\label{10-35}%
\nu^2:=(c/l)^2\simeq c^2\Lambda_+/3\sim
10^{-35}sec^{-2},%
\ee%
where $\nu:=c/l$ is called the Newton-Hooke constant. Thus, their
effects can still be ignored up to now locally
  except for the cosmic scale physics.

  The paper is arranged as follows.  In sec. \ref{sec: UWFH}, we
 briefly recall  the inertial motion group based on
 the \PoRcl\ and the relativistic kinematics that with common Lorentz isotropy
  act naturally as the \SR\
 triple with four related  algebras.
 In sec. \ref{sec: P-double}, we show that
 there is the dual  Poincar\'e group  ${\cal P}_2$ and its kinematics on a pair of degenerate Einstein manifolds
$M_\pm$ induced from the origin lightcone $C_O$ as absolute for
its space/time-like region $R_\pm$, respectively. Thus, there is
the Poincar\'e double $[{\cal P}, {\cal P}_2]_{M/M_\pm}$ consists
of the Poincar\'e group ${\cal P}$ on \Mink-space and the  ${\cal
P}_2$-kinematics on $M_\pm$ for $R_\pm$, respectively. In sec.
\ref{sec: SRT}, we show that
with common Lorentz isotropy not only the \SR\ on %two
 Beltrami-\dS/\AdS\ spaces are  related, but their domain conditions and  absolutes
 at origin  also become the %space/time-like region
$R_\pm$. This can also be reached from the
 shifted \Mink-lightcone
by relaxing the flatness of  \Mink-space. %so that there is the \dS/\AdS\ double. We also show that
Hence, there are naturally six doubles that form the quadruple
${\frak Q}_{PoR}$ for the \SR\ triple.
 Finally, we  end with
some remarks.

%%%%%%%%%%%%%%%%%%%%%%%%%%%%%%%%%%%%%%%%%%%%%%%%%%%%%%%%%%
%%%% PoRcl and IM(4)
%%%%%%%%%%%%%%%%%%%%%%%%%%%%%%%%%%%%%%%%%%%%%%%%%%%%%%%%%%

\section{Principle of Relativity with Two Universal Constant and  Inertial Motion Group}\label{sec: UWFH}
As was just emphasized, in order to explore whether there is new
kinematics for the cosmic scale physics,  the \PoR\ and the
postulate for the light propagation in Einstein's \SR\ \cite{1905}
should be extended and weakened to the \PoR\ of two universal
constants $(c,l)$, i.e. the \PoRcl. Let us consider kinematic
aspects of the \PoRcl. That is what should be the symmetry for the
inertial motions with respect to  the \PoRcl. It turns out there is
a group for inertial motions described by Newton's law of inertia
called the inertial motion group, which is {homomorphic} to 4d real
projective group $PGL(5,R)$, and all kinematics with space isotropy
of ten generators with some missed before.

\subsection{Inertial Motion Transformation Group for Newton's Law of Inertia}

Umov\cite{Umov} , Weyl\cite{Weyl}, Fock\cite{Fock} and Hua\cite{Hua,
HuaC} studied an important issue: What are the most general
transformations among the inertial frames ${\cal F}:=\{S(x)\}$ that
keep the inertial motions?

In view the \PoRcl, the answer %(see references in \cite{SRT, SRT2})
 is: {\it The most general
transformations among ${\cal F}$ that keep
 inertial
motion described by Newton's first law
 \be\label{eq:uvm}%
 x^i=x_0^i+v^i(t-t_0),~~
v^i=\frac{dx^i}{dt}={\rm consts}.~~ i=1, 2, 3, %
\ee%
are  the  \LFT s    of twenty-four parameters %
\be\label{eq:LFT} %
{\rm T}:\quad l^{-1}x'^\mu = \frac{A^\mu_{\ \nu} l^{-1} x^\nu +
b^\mu}{c_\lambda l^{-1}x^\lambda + d},\quad x^0=ct,
\ee%
\be\label{eq:det}
  det\ {\rm T}=\left| \begin{array}{rrcrr}
    A    & b^t \\
    c%_0 & c_1 & \ldots & c_n
     & d
  \end{array} \right|
  \neq 0,
\ee%
where $A=(A^\mu _{~\nu})$  a $4\times 4$ matrix, $b,c$ $1\times 4$
 matrixes, $d\in R$, $c_\lambda x^\lambda=\eta_{\lambda\sigma}c^\lambda
 x^\sigma$,
  $\small{^ t}$
 for  transpose and  $J=(\eta_{\mu\nu})=diag(1,
  -1,-1,-1)$. These $LFTs$
  form the inertial motion
 group $IM(1,3)$ homomorphic to  the 4d real projective
group ${PGL}(5,R)$, i.e. $\forall {\rm T}\in IM(1,3)\sim
PGL(5,R)$, with the inertial motion algebra $\mathfrak{im}(1,3)$
isomorphic to $\frak{pgl}(5,R)$.} Further, the time reversal ${\bf
T}$ and space inversion ${\bf P}$ preserve the inertial motion
(\ref{eq:uvm}). So, all  issues are modulo the $\bf T$ and ${\bf
P}$ invariance.

In fact, inertial motion (\ref{eq:uvm}) can be viewed as a 4d
straightline \cite{Hua, HuaC}.  In  projective geometry, \LFT s
(\ref{eq:LFT}) form the real projective group ${PGL}(5,R)$\cite{Hua}
with algebra $\frak{pgl}(5,R)$. But, for orientation in physics the
antipodal identification of 4d real projective space should not be
taken so that $IM(1,3)\thicksim{PGL}(5,R)$. Hereafter, we still call
it the projective geometry approach in this sense.

\subsection{Inertial Motion Algebra and Relativistic
Kinematic Algebras}
 It is straightforward to get the generator set
 $\{T\}^\mathfrak{im}:=(\H^\pm, \P^\pm_i,\J_i,\K_i,\N_i,\R_{ij}, M_\mu)$ of
  \LFT s (\ref{eq:LFT})
where
 \be\nno%
%\begin{split}
 &\N_i := t \partial_i + c^{-2} x_i
\partial_{t},&\\\label{eq:R,M}
&R_{ij}:= x_i \partial_j + x_j \partial_i, (i< j),\,\, M_\mu :=
x^{(\mu)} \r_{(\mu)},&
%\end{split}
\ee%
where no summation %is taken
for  repeated  indexes in brackets. {It is important that either
$\frak{so}(1,3)$ of Lorentz group
${\cal L}$ generated by  space rotation ${\bf J}_i$ defined as%
  \be\label{eq:J}%
 \J_i=\frac{1}{2}\epsilon_i^{\,jk}L_{jk}, \ \, L_{jk}:=x_j\partial_k-x_k\partial_j,%
 \ee%
  and Lorentz
boosts $\K_i$ defined as%
\be\label{eq:Ki}%
 \K_i:=t \partial_i -c^{-2} x_i
 \partial_t,%
\ee%
or $\frak{so}(3)$ of its subgroup $SO(3)$ generated
by space
  rotation  ${\bf J}_i$,
 both groups are
  subset of $A^\mu_{\ \nu}\subset T$ in (\ref{eq:LFT}). So the Lorentz isotropy for
  relativistic cases or  space isotropy for all cases are common for relevant kinematics, respectively.}  Among four  time and space
translations $\{{\cal H}\}:=\{ \H, \H', H^\pm\}$ of dimension
$[\nu]$,  $\{{\bf P}\}:=\{ {\bf P}_i, {\bf P}'_i, {\bf P}^\pm_i\}$
of dimension $[l^{-1}]$, and four boosts $\{ \K\}:=\{\K_i, \N_i,
 \K^{\mathfrak{g}}_i,\K^{\mathfrak{c}}_i\}$ of Lorentz, % one ${\bf K}_i$,
 geometry, % one ${\N}_i$,
 Galilei %one ${\bf K}_i^g$
  and Carroll boosts of dimension $[c^{-1}]$,  there are two
independents, respectively
  \be \label{eq:H}%\nno%[1mm]
&H:=\partial_t,\,\, H':=-\nu^{2}t x^\nu\partial_\nu,\,\, H^\pm
:=\partial_t\mp \nu^{2}t x^\nu\partial_\nu;&\\\label{eq:P}%\nno%
 &\P_i:=\partial_i,\, \P'_i:=- l^{-2}x_i
 x^\nu\partial_\nu,\, {\mathbf P}^\pm_i :=\partial_i\mp l^{-2}x_i
 x^\nu\partial_\nu;&\\\nno%
& \K_i:=t \partial_i -c^{-2} x_i
 \partial_t,\,\,\, \N_i:=t \partial_i +c^{-2} x_i
 \partial_t,&\\\label{eq:K}& \K_i^\frak{g}:=t \partial_i,\quad \K_i^\frak{c}:= -c^{-2} x_i
 \partial_t.
\ee%
These generators are scalar and vector representation of
$\frak{so}(3)$ generated by  ${\bf J}_i$ without dimension, as
follows\cite{SRT, SRT2}
\be\label{eq: JHPK}%\nno%
[ \J,\J ]=\J, \,\,[\J, {\cal H}]=0,\,\, [\J, {\P}]={\P},
[\J, {\K}]={\K},%\\\nno%[1mm]%
\ee%
where with
$\epsilon_{123}=-\eps_{12}^{\ \ 3}=1$ % $x_\mu=\eta_{\mu\nu}x^\nu$ %i.e. %the Newton-Hooke const. $\nu:=c/l$, $x_0=x^0, x_i=-\delta_{ij}x^j$,
 $[\J,\P]=\P$ is, e.g. a shorthand % abbreviation
of $[\J_i,\P^\pm_j]=-\epsilon_{ij}^{~~k}\P^\pm_k$ etc.  All
generators and commutators  have right dimensions expressed by the
constants $c, l$ or $\nu$. In addition to combination
(\ref{eq:combinatory}), there are also combinatory relations
$\K_j/\N_j=\K_i^\frak{g}\pm \K_i^\frak{c}$ for boosts. %

Then the generator set
 $\{T\}^\mathfrak{im}$ spans the
 $\frak{im}(1,3)$ as follows
 \cite{SRT2},
 \begin{eqnarray}
\begin{array}{l}
\, {[}\P^+_i, \P^-_j{]} =(1-\delta_{(i)(j)})l^{-2} R_{(i)(j)} -
2l^{-2}
\delta_{i(j)} (\M_{(j)}+D), \\
\begin{array}{ll}
{[}\P^\pm_i, \M_j{]}= \delta_{i(j)}\P^\mp_{(j)},&
[\P^{\pm}_i, \R_{jk}]=-\delta_{ij}\P^{\mp}_k-\delta_{ik}\P^{\mp}_j,\\
{[}\H^+, \H^-{]} =  2{ \nu^2} \left(\M_0 + D\right), &
[\H^\pm, \M_0]= \H^\mp,  \\
{[}\K_i, \M_0{]}=-\N_i, &   [\K_i, \M_j]= \delta_{i(j)} \N_{(j)},\\
{[}\K_i, \R_{jk}{]}= -\delta_{ij} \N_k-\delta_{ik}\N_j,& [\N_i, \M_0]= -\K_i,\\
{[}\N_i, \M_j{]}= \delta_{i(j)}\K_{(j)},&
{[}\N_i,\R_{jk}{]}=-\delta_{ij} \K_k-\delta_{ik} \K_j, \\
{[}\K_i,\, \N_j{]} =(\delta_{(i)(j)}-1)  c^{-2} R_{(i)(j)} -&
2\delta_{(i)j}c^{-2}\left(M_{0} - M_{(i)}\right),\\
{[}L_{ij}, \M_k{]}=\delta_{j(k)}  R_{i(k)}-\delta_{i(k)}R_{j(k)}
,& {[}R_{ij}, \M_k{]}=\delta_{i(k)}L_{j(k)} +
\delta_{j(k)}L_{i(k)},
\end{array}\\
\,
{[}L_{ij},\R_{kl}{]}=2(\delta_{ik}\delta_{jl}+\delta_{il}\delta_{jk})\,
(\M_i  - \M_j
) + \delta_{ik}\R_{jl} +\delta_{il}\R_{jk}-\delta_{jk} R_{il}- \delta_{jl}R_{ik}, \\
\, {[}\R_{ij},R_{kl}{]} = -
\delta_{ik}L_{jl}-\delta_{il}L_{jk}-\delta_{jk}L_{il}-\delta_{jl}L_{ik},
%\,{[}\P^\pm, M_0]=[H^\pm, M_i{]}=[\H^\pm,
%R_{ij}]={[}L_{ij},\M_0{]}={[}R_{ij}, \M_0{]}=[M_i,M_j]=0.
\end{array} \label{im4}
\end{eqnarray}%
where $D$ is called the dilation generator  and is defined as%
\be\label{eq:D}%
D:=\sum_\ka M_\ka.%
\ee%

In Table I, all relativistic kinematics are listed {symbolically.
For the geometrical and non-relativistic cases, we shall study
them in detail elsewhere.} {\bf\begin{table}[thp] \caption{ All
relativistic
 kinematics}\vskip 2mm {\scriptsize
\begin{tabular}{|c c c c c c c c|}
\hline Group & Algebra & Generator set
  & $[\cal H,\P]$ & $[\cal H,\K]$ & $[\P,\P]$ &$[\K,\K]$ &$[\P,\K]$\\
\hline
$\begin{array}{c}{\cal P}\\{\cal P}_2\end{array}$ &$\begin{array}{c} \mathfrak{p}\\
\mathfrak{p}_2\end{array}
$ & $\begin{array}{c}(H, \P_i, \K_i, \J_i)\\
\{H', \P'_i, \K_i, \J_i\}\end{array}$ & 0 & $\P$ & 0
&$-c^{-2}\J$&$c^{-2}\cal H$\\${\cal D}_+$  &$\mathfrak{d}_+$ &
$(H^+, \P^+_i, \K_i, \J_i)$ & $\nu^2\K$ & $ \P$ & $l^{-2}\J$ &
$-c^{-2}\J$  &$c^{-2}\cal H$\\
${\cal D}_-$ & $\mathfrak{d}_-$ &  $(H^-, \P^-_i, \K_i, \J_i)$ &
$-\nu^2\K $ & $ \P$ & $-l^{-2}\J $
& $-c^{-2}\J$ & $c^{-2}\cal H$\\
\hline
\end{tabular}
}

\end{table}}

It is clear that
with common Lorentz isotropy $\frak{so}(1, 3)$ spanned by   $(\K_i, \J_i)$, the Poincar\'e %algebraic
doublet $(\frak{p},\frak{p}_2)$ leads to  the \dS/\AdS\ %algebraic
doublet $(\frak{d}_+,\frak{d}_-)$  with the Beltrami time and
space translations $(\H^\pm, \P^\pm_j)$ in the Beltrami-\dS/\AdS-space, respectively, and vice
 versa\cite{SRT, SRT2}. In addition,
from the algebraic relations  of $\frak{im}(1,3)$ \cite{SRT2} both
$(\frak{p},\frak{p}_2)$ and $(\frak{d}_+,\frak{d}_-)$ are closed
in the  $\frak{im}(1,3)$, while the  generators
 $(R_{ij}, M_\mu)$, which generate $A_{\,\,\nu}^\mu$ in \LFT s (\ref{eq:LFT}) together
 with $\N_i$ in (\ref{eq:K}) and
 $(\K_i, \J_i)$, % of Lorentz algebra $\frak{so}(1,3)$,
 exchange the  translations
  from one to another in  {$(\frak{p},\frak{p}_2)$ and $(\frak{d}_+,\frak{d}_-)$.
   This has been shown in \cite{SRT,SRT2}.} Since all issues are in  $IM(1,3)$ based on the
\PoRcl,  there should be
three kinds of \SR\cite{Lu,LZG,1950s,BdS,Duality} { with Newton's law of inertia} %
that {automatically} form {the \SR\ triple} of four related
algebras {${\frak so}(1,3)=\frak{p}\cap \frak{p}_2
\cap\frak{d}_+\cap\frak{d}_-$} that form the algebraic quadruplet
$\frak{q}$\cite{SRT2}.

We will show that this also the case for their relevant groups and geometries. %

%%%%%%%%%%%%%%%%%%%%%%%%%%%%%%%%%%%%%%%%%
%  The Poincare double
%%%%%%%%%%%%%%%%%%%%%%%%%%%%%%%%%%%%%%%%%%
\section{Dual  Poincar\'e Group, ${\cal P}_2$-Kinematics and Poincar\'e Double}\label{sec: P-double}
Let us now consider why together with usual Poincar\'e group there
is a dual  Poincar\'e group and its role in relativistic physics.

\subsection{Poincar\'e Group and Dual  Poincar\'e Group}

 Under usual Poincar\'e group
 ${\cal P}:=ISO(1,3)=R(1,3)\ltimes \cal L$ %transforms %o arbitrary dilations%, but there are no restricts on other 14 parameters. %which keep Einstein's light-cone to be invariant at any given point.
\be\label{eq:ISO(1,3)}%
 {P}: \,\,
 %x^\mu \to
 x'^\mu=
 (x^\nu-a^\mu)L^\mu_{\ \nu},\,\, det{P}= det \left (
\begin{array}{rrcrr}
    L & b^t \\
    0 & 1
  \end{array} \right)
  \neq 0,%
\ee%
where $L=(L^\mu_{\ \nu})\in  SO(1,3)$, $ b^t=-l^{-1}(aL)^t$,  and
$\forall P\in {\cal P}$. %The universal constant $l$ is hidden in Einstein's \SR.
Then, it follows the \Mink-space $M:={\cal P}/{\cal
L}$ with  the \Mink-metric and the \Mink-lightcone
%with the apex
at  event $A(a^\mu)$ %
\be\nno%
&ds^2=\eta_{\mu\nu}dx^\mu
dx^\nu=dxJdx^t,&\\\label{eq:MconeatA}
&C_A:\,\,\,% \eta_{\mu\nu}(x^\mu-a^\mu) (x^\nu-a^\nu)=
(x-a)J(x-a)^t=0,&  % \gtreqless 0,\qquad
\ee%
 with the lightcone structure $[C_A]$ %with \Mink-metric%  the %following
:%
 \be\label{eq:CA}
[C_A]:\,\, (x-a)J(x-a)^t \gtreqless0,\,\,ds^2|_A=dxJdx^t|_A=0.%\eta_{\mu\nu}x^\mu x^\nu\gtreqless0~\Leftrightarrow~\eta_{\mu\nu}x'^\mu x'^\nu\gtreqless0.%
\ee%
The generator set $\{T\}^{\mathfrak{p}}:=(H,\P_i, \K_i,\J_i)$  spans %the
$\frak{p}:=\mathfrak{iso}(1,3)$ listed in Tale I.
%Poincar\'e algebra%
%listed in Table I symbolically.

It is important that based on the \PoRcl, in addition to this usual
Poincar\'e group ${\cal P}$, there is another group isomorphic to it
called the dual  Poincar\'e group ${\cal P}_2\cong ISO(1,3)$. Let us
now consider this group and its kinematics.

In fact, the symmetry of the origin
lightcone $C_O$ with its space/time-like region $R_\pm$%
\be\label{eq:CO}%
&C_O: \quad xJx^t=0&\\\label{eq:Rpm}%
&R_\pm: \quad r_\pm(x):=\mp xJx^t> 0& \ee%
 is not just the Lorentz group,
but the semi-product of the dual  Poincar\'e group ${\cal P}_2$ and
a dilation generated by (\ref{eq:D}).

 By means of the \PGL\ method,
 $C_O$ and  $R_\pm$ should be regarded as the absolute and the domain,
  respectively. Then, \LFT s  (\ref{eq:LFT}) reduce to their
   subset: %with $A^\mu_{\ \nu}=L^\mu_{\ \nu} \in SO(1,3), b^\mu=0$ and $d=1$:
\be\label{eq:LFTP2}%\nno%
 l^{-1}x^\mu\to  l^{-1}x'^\mu = \frac{L^\mu_{\ \nu}l^{-1} x^\nu }{c_\lambda l^{-1} x^\lambda +
  d}, %\,\,~ c_\lambda l^{-1} x^\lambda +d\neq 0,%
  \ee%
in which the \LFT s for $d=1$ form the dual
Poincar\'e group, $\forall P_2\in {\cal P}_2 \cong ISO(1,3)$, %, i.e.%
\be\label{eq:P2}%
 det P_2= det \left ( \begin{array}{rrcrr}
    L & 0 \\%^0_{\ 0} & A^0_{\ 1} & \ldots & A^0_{\ n} & b^0 \\
    c & d
  \end{array} \right)
  \neq 0,\,\,\, {\rm for}\,\,\, d=1.&
\ee%
This can be easily seen from that each matrix $P_2$ is isomorphic
to the transport of a matrix $P$ in (\ref{eq:ISO(1,3)}). It is
also easy to show that this group preserves $C_O$ and $R_\pm$ so
does $[C_O]$  in (\ref{eq:CA}) with $A(a^\mu)=O(0^\mu)$. The
generator of dilation  $d$ in (\ref{eq:LFTP2}) is $D$
(\ref{eq:D}).

The generator set $\{T\}^{\frak{p}_2}=
 (\H',
 \P'_i,\K_i,\J_i)$ of ${\cal P}_2$ consists of %
 \be\nno%
& H':=-\nu^{2}t x^\nu\partial_\nu,\ \,\P'_i:=- l^{-2}x_i
 x^\nu\partial_\nu,&\\\label{eq:p2set}%
 & \K_i:=t \partial_i -c^{-2} x_i
 \partial_t,\ \ \J_i=\frac{1}{2}\epsilon_i^{\,jk}L_{jk}, &
 \ee%
 spans
 $\frak{p}_2 \cong\frak{iso}(1,3)$ for $[C_O]$
  listed in Table I.  The
generator $D$ for the dilation  also keeps $[C_O]$.

In fact, the dual  Poincar\'e group  is also based on the \PoR.
However, it had been simply reduced
to the homogeneous Lorentz group in \cite{1905}, in which Einstein wrote:%\cite{1905}: % in 1905.
\bc
\begin{minipage}{160mm}
\it At the time $t=\tau=0$, when the origin of the co-ordinates is
common to the two frames, let a spherical wave be emitted therefrom,
and be propagated with the velocity $c$ in system $K$. If $(x, y,
z)$ be a point just attained by this wave, then
$$x^2+y^2+z^2=c^2t^2.$$
 Transforming this equation with the aid of
our equations of transformation we obtain after a simple calculation
$$\xi^2+\eta^2+\zeta^2=c^2\tau^2.$$
The wave under consideration is therefore no less a spherical wave
with velocity of propagation $c$ when viewed in the moving system.
This shows that our two fundamental principles are compatible
\footnote{The equations of the Lorentz transformation may be more
simply deduced directly from the condition that in virtue of those
equations the relation $x^2+y^2+z^2=c^2t^2$ shall have as its
consequence the second relation $\xi^2+\eta^2+\zeta^2=c^2\tau^2$.}.
 \end{minipage}\ec

In the above footnote by Einstein, $H'$ and $\P'_i$ had been missed.

 As is well-known, for the light cone, the maximum invariant
symmetry is conformal group $SO(2,4)$. However, its special
conformal transformations do not preserve the inertial motion
(\ref{eq:uvm}). It should also be mentioned that for $L^\mu_{\
\nu}=\delta^\mu_{\ \nu}$, transformations (\ref{eq:LFTP2}) look
like  ``local scale transformations" at first glance. In fact,
this is not true. The local scale transformations should not
include other group parameters $c_\lambda$ and $d$, which do not
depend on $x$.

\subsection{Dual  Poincar\'e Group as Kinematics on Degenerate Einstein Manifolds }

As a kinematic symmetry, what kind of spacetimes should be
transformed under the dual  Poincar\'e group? Namely, what about the
dual  Poincar\'e group ${\cal P}_2$ invariant kinematics?

It can be show that the ${\cal P}_2$ invariant 4d metric must be
degenerate\cite{zhou, Huang0909}. We will study this issue
together others by means of the \PGL\ method as well.

For a pair of events $A(a^\mu), B(b^\mu)\in R_\pm$,  a line
between them%
\be\label{eq:line}%
L:\quad (1-\tau)a+\tau b%
\ee%
 crosses  ${C}_O:=\partial(R_\pm)$ at $\tau_1$ and
$\tau_2$, which satisfy%
\be%
(b - a) J (b-a)^t \tau^2 + 2 a J (b - a)^t + (aJa^t
 \mp l^2)  = 0. \ee%
 For four events with $\tau=(\tau_1, 1,
\tau_2, 0)$, a cross ratio
can be given. From the power 2 cross ratio invariant%
\be\label{Deltapm}%
&{\Delta}_{R_\pm}^2(A, X)
=\pm l^{2}\{r_\pm^{-1}(a)r_\pm^{-1}(x)r_\pm^2(a,x)-1\},&\\\nno% \quad
&r_\pm(a):=r_\pm(a,a)>0,\,\ r_\pm(a,x):=\mp aJx^t,& %
\ee%
 for
$X(x^\mu), X+dX(x^\mu+dx^\mu)\in R_\pm$, it follows a degenerate
metric
\be\nno%
&{g}_{\pm \mu\nu}= %l^{-2}
( \frac{\eta_{\mu\nu}}{r_\pm(x)}\pm
 l^{-2}\frac{x_\mu
x_\nu}{r^2_\pm(x)}),\,\ %dx^\mu dx^\nu,
\,x_\mu=\eta_{\mu\lambda}x^\lambda,&\\ \label{gpm}%
&{\vect g}_\pm:=({ g}_\pm)_{\mu\nu}=%\d {l^{-2}}
{r^{-1}_\pm(x)} ( J \pm l^{-2} \d {J x^t x J }{r_\pm(x)}).
\ee%
%and other invariant geometric objects follow, respectively.
Although metric (\ref{gpm}) is degenerate, i.e. $det {\vect
g}_\pm=0$, formally there is still a contra-variant metric as its
inverse,
respectively: %
\be\label{hpm}%
&{\vect g}^{-1}_\pm= %{l^{2}}
{r_\pm(x)} ( J
\pm  l^{-2}\d {J x^t x J }{r_\pm(x)})^{-1},&%
\ee%
which is divergent. But, their Christoffel symbol can
still be formally calculated. This meaningfully results% reads%
 \be\label{Chrs}%
 \Gamma^\lambda_{\pm
\mu\nu} (x)= \pm r_{\pm}^{-1}(x) ( \delta^\lambda_\mu x_\nu +
\delta^\lambda_\nu x_\mu),%
 \ee%
which is obviously metric compatible, i.e.%
\be\label{eq:g;=0}%
\nabla_{\Gamma_\pm}{\vect g}_\pm=0,
\,\,\,\,\nabla_{\Gamma_\pm}{\vect
g}^{-1}_\pm=0. \ee%
And it is straightforward  to get the Riemann, %curvature,
Ricci
 %curvature
  and scale curvature, respectively, as follows
 \be
&R^\mu_{\pm \nu \la \si}(x)=\pm
l^{-2}(g_{\pm\nu\la}\dl^\mu_\si-g_{\pm\nu\si}\dl^\mu_\la)&\\\label{riccicur}
&R_{\pm \mu \nu }(x)%= R^\la_{\pm \mu \nu \la}
= \pm 3 l^{-2}g^\pm_{\mu\nu}, \,~~\,R_\pm(x) = \pm  12 l^{-2}.&%
\ee %
Then the  $M_\pm:=(M^{{\cal P}_2}_\pm, {\vect g}_\pm,
{\vect g}^{-1}_\pm, { \nabla_{\Gamma_\pm}})$ %with degenerate metric  (\ref{gpm}) and its inverse (\ref{hpm})
is an Einstein
 manifold with %cosmological constant
 $\Lambda_\pm=\pm 3/l^2$ for $R_\pm$ (\ref{eq:Rpm}), respectively. It easy to check
 that the Lie derivatives of these objects vanish with respect to  the ${\mathfrak p}_2$-generators
 as Killing vectors on  $M_\pm$,
 respectively.
 And  Eq. (\ref{eq:uvm}) is indeed the
  equation of motion
 for a free particle, if any, on such a pair of Einstein manifolds
 $M_\pm$. Namely, there is also Newton's first law of inertia.

 Since all these  objects are given originally from
 the cross ratio invariance of $C_O$ (\ref{eq:CO}) under \LFT s (\ref{eq:LFT}) based on the \PoRcl,
 such a ${\cal P}_2$-kinematics is
 invariant under  (\ref{eq:P2}).
  Another  ${\cal P}_2$-degenerate geometry for $R_+$
 is given in a different  manner \cite{Huang0909}. % without the \PoRcl.

\subsection{The Poincar\'e Double}

In fact, the dual  Poincar\'e group ${\cal P}_2$ also exists  for lightcone structure  $[C_A]$ (\ref{eq:CA})  %not only for the origin lightcone, but also
while  the ${\cal P}_2$ for  $[C_O]$ (\ref{eq:CO})  is a
representative among them. Then there are intersections for ${\cal
P}$ and ${\cal P}_2$ as well as for $M$ and $M_\pm$, {i.e. ${\cal
L}= {\cal P}\cap{\cal P}_2$ and $R_\pm = M\cap M_\pm$,}
respectively. Thus, related to the \Mink-space there exist
infinite many Poincar\'e doubles with  the Poincar\'e double
$[{\cal P},{\cal P}_2]_{M/M_\pm}$ at the origin as a representative for  $M$ and
a pair of %Einstein manifolds
$M_\pm$ induced from $C_O$ (\ref{eq:CO})
 for  $R_\pm$ (\ref{eq:Rpm}), respectively.

 Actually, under transformations
 (\ref{eq:ISO(1,3)}) of  $\cal P$, the
 Poincar\'e algebraic doublet $(\frak{p},\frak{p}_2)$ at origin $O(o^\mu)$ can be
transformed to $A(a^\mu)$ and vice versa. In fact, under $\cal P$,
the generators $\P'_\mu:=(H', \P'_i)$ as a 4-vector on $M$
 are transferred and the action is closed in the  algebra
 $\frak{im}(1,3)$
 \be\label{eq:[p,p']}%
 {\cal L}_{\P_\mu} \P'_\nu =[\P_\mu,\P'_\nu]\omits{=-l^{\-2}\eta_{\mu\nu}\sum_{\lambda}M_\lambda
 -l^{-2}\frac{1}{2}R_{\mu\nu}} \in \frak{im}(1,3)\cong
 \frak{pgl}(5,R),
 \ee%
in which there are \dS/\AdS\ algebras $\frak{d}_\pm$ for \dS/\AdS\
\SR\ as subalgebras.

It should be emphasized that the existence of the dual Poincar\'e
symmetry is indicated by a theorem in projective geometry: In the
group $PGL(2,K)$ of 1d projective space over field $K$, the
invariant group of  any given point is isomorphic to the affine
subgroup\cite{HuaWan}{\footnote{The $PR^1$ is a 1d compact and
differentiable manifold with its transformation group $PGL(2,R)$.
There are at least two inhomogeneous coordinate patches to cover it.
In each of them \LFT s (\ref{eq:LFT}) may transform a point to its
 infinite, which must be included so the patch is extended to $PR^1$. And in each
  of such extended patches the theorem holds. In the
intersection, the  affine subgroups in one extended patch is just
the dual  ones in the other.}}. This theorem {can  also be extended
to } $PGL(5,R)$ of 4d real projective space. If the Euclidian metric
$\delta_{\mu\nu}$ related to the 4d affine space is changed to
$\eta_{\mu\nu}$, the fixed point is then changed to the \Mink\
lightcone and the dual  Poincar\'e group appears.

Thus,  symmetry and geometry related to the \Mink-space are
dramatically changed. In order to eliminate the double and restore
Einstein's \SR, $\l = \infty$ should be taken. But,
 the
cosmological constant $\Lambda_+$ leads to  $[{\cal P},{\cal
P}_2]_{M/M_+}$ with  bound (\ref{10-35})  for  modern relativistic
physics.

%%%%%%%%%%%%%%%%%%%%%%%%%%%%%%%%%%%%%%%%%%%%%%%%%%%%%%%%%%%%%%%%%%%%%%%%
%%%%%%%%%%%%%%
%%%%%%%%%%%%%%%%%%%%%%%%%%%%%%%%%%%%%%%%%%%%%%%%%%%%%%%%%%%%%%%%%%%%%%
\section{Relativistic Quadruple with Common Lorentz
Isotropy}\label{sec: SRT}

As was shown in \cite{SRT, SRT2}, dual to the Poincar\'e algebraic
doublet $(\frak{p},\frak{p}_2)$  at the origin, there is a \dS/\AdS\
algebraic doublet $(\frak{d}_+,\frak{d}_-)$   at the same origin.
And together other four doublets   at the origin, there is a
relativistic algebraic quadruplet ${\frak
q}=(\frak{p},\frak{p}_2,\frak{d}_+,\frak{d}_-)$   at the origin,
which is in fact a representative of infinite many such kind of
quadruplets. We will show this is also the case for group and
geometry aspects.

\subsection{De Sitter, Anti-de Sitter  Relativistic Kinematics and De Sitter Double}

First, %from the \Mink-lightcone,
corresponding to the algebraic doublet $(\frak{d}_+,\frak{d}_-)$
with common Lorentz isotropy algebra $\frak{so}(1,3)$, there is the
\dS/\AdS\ double $[{\cal D}_+,{\cal D}_-]_{D_\pm}$ of common Lorentz
isotropy group ${\cal L}\subset IM(1,3)$   at the origin.
%and relations among  four groups ${\cal P},{\cal P}_2$, ${\cal D}_+$ and ${\cal D}_-$
%This  can also be reached   by relaxing the flatness of
%\Mink-space.

 In fact, in terms of homogeneous projective coordinates  the  \dS/\AdS-hyperboloid
  ${\cal H}_\pm$ and their
boundaries can be expressed,
respectively%
\be\label{eq:hyperboloids}%
 &{\cal H}_\pm:\quad \eta_{\mu \nu}\xi^\mu \xi^\nu \mp (\xi^4)^2 \lessgtr 0,&\\\label{eq:BdyH}%
&{\cal B}_\pm=\partial{\cal H}_\pm:\quad \eta_{\mu \nu}\xi^\mu
\xi^\nu \mp
 (\xi^4)^2=0.&
\ee%
It is clear that they are invariant under \dS/\AdS\ group, i.e.
${\cal D}_\pm:=SO(1,4)/SO(2,3)$, and the $\eta_{\mu \nu}\xi^\mu
\xi^\nu$ and $(\xi^4)^2$ are in intersections of the ${\cal H}_\pm$
and their boundaries ${\cal B}_\pm=\partial{\cal H}_\pm$,
respectively. These imply that they just share the common Lorentz
isotropy as required.

In terms  of the Beltrami coordinates as  the inhomogeneous
projective
coordinates without antipodal identification, i.e. $x^\mu$ in a chart $U_4$, say, %of the Beltrami atlas
\be\label{eq:BLxU4}%
x^\mu=l{\xi^4}^{-1}{\xi^\mu},\quad \xi^4> 0,%
\ee%
the   \dS/\AdS-hyperboloid
  ${\cal H}_\pm$ (\ref{eq:hyperboloids}) and their
boundaries (\ref{eq:BdyH}) become  the domain conditions and
absolutes for the Beltrami-\dS/\AdS-space of
 radius $l$ with common Lorentz isotropy, respectively%
\be\label{eq:domains}%
&\frak{D}_\pm:\quad \sigma_\pm(x):=\sigma_\pm(x, x)=1\mp l^{-2}xJ
x^t
>
 0,&\\%
\label{eq: sigma=0}%
&\frak{B}_\pm:\quad  \sigma_\pm(x)=1\mp l^{-2}xJ x^t = 0.&%
 \ee%

 Then, by means of the \PGL\ method, the intersected
Beltrami-\dS/\AdS-spaces $D_\pm$ of ${\cal D}_\pm$ invariant  can
be set up  and form the \dS/\AdS\ double $[{\cal D}_+,{\cal
D}_-]_{D_\pm}$ with common Lorentz isotopy, i.e. $\cal L={\cal
D}_+\cap{\cal D}_-$.

In fact,  \LFT s (\ref{eq:LFT}) with (\ref{eq: sigma=0}) as
absolute are reduced to  the \dS/\AdS-\LFT s with  common
$L_{~\ka}^\mu\in\cal L$%={\cal D}_+\bigcap{\cal D}_-$
\be\nno%
&{\cal D}_\pm:~ x^\mu\rightarrow {x'}^\mu=\pm
\sigma_\pm^{1/2}(a)\sigma_\pm^{-1}(a,x)(x^\nu-a^\nu)D_{\pm~\nu}^{~\mu},&\\\label{eq:SO14,23}%\nno
&D_{\pm~\nu}^{~\mu}=L_{~\nu}^\mu\pm l^{-2}%
a_\nu a^\ka (\sigma_\pm(a)+\sigma_\pm^{1/2}(a))^{-1}L_{~\ka}^\mu.&
\ee
This is the same as given before in \cite{Lu,LZG,BdS}. As  the \LFT s of ${\cal D}_\pm$, (\ref{eq:SO14,23}) %which transform a point $A(a^{\mu})$ %with $\sigma_\pm(a^{\mu})>0$
%in  $S(x)$  to the origin of  $ S'(x')$ and
preserves  %domain
 (\ref{eq:domains})  for the Beltrami-\dS/\AdS\ space $D_\pm$, respectively.

For a pair of events $A(a^\mu), B(b^\mu)\in \frak{D}_\pm$,  a line
(\ref{eq:line})
between them%
 crosses  the absolutes $\frak{B}_\pm$ %${C}_O:=\partial(R_\pm)$
  at $\tau_1$,
$\tau_2$. For four events with $\tau=(\tau_1, 1, \tau_2, 0)$, a
cross ratio can be given. Further, from a power 2 cross ratio
invariant the following interval between a pair of events
$A(a^\mu)$ and $X(x^\mu)$
 % Beltrami metric
  and the
lightcone with top at $A(a^\mu)$  follows, respectively
 \be\label{Dlt}%
 &{\Delta}_\pm^2(A,
X) =\pm
l^{-2}\{\sigma_\pm^{-1}(a)\sigma_\pm^{-1}(x)\sigma_\pm^2(a,x)-1\}\gtreqqless
0,
&\\\label{eq:Blightcone} %
&{\cal F}_{\pm}: \quad
\sigma_\pm^2(a,x) -\sigma_\pm(a)\sigma_\pm(x)=0.&% 
\ee %
For the  closely nearby two events $X(x^\mu),
X+dX(x^\mu+dx^\mu)\in \frak D_\pm$,  the Beltrami
metric\cite{Lu,LZG,BdS} follows from (\ref{Dlt})
\be\label{eq: BKmetric}%
&ds_\pm^2=(\frac{\eta_{\mu\nu}}{\sigma_\pm(x)}\pm l^{-2}\frac{
x_\mu x_\nu}{
\sigma_\pm^{2}(x)})dx^\mu dx^\nu,\,\,\,\sigma_\pm(x)>0.&%
\ee%
That is%
\be \label{gbpm}%
&{\vect g}^B_\pm:=({ g}^B_\pm)_{\mu\nu}=%\d {l^{-2}}
{\sigma^{-1}_\pm(x)} ( J \pm l^{-2} \d {J x^t x J
}{\sigma_\pm(x)}).
\ee%
Its inverse as the contravariant metric reads%
\be\label{hbpm}%
&{\vect g}^{B-1}_\pm= %{l^{2}}
{\sigma_\pm(x)} ( J
\pm  l^{-2}\d {J x^t x J }{\sigma_\pm(x)})^{-1}.&%
\ee%

Due to transitivity of (\ref{eq:SO14,23}), the Beltrami-\dS/\AdS\
space ${D}_\pm \cong {\cal D}_\pm/\cal L$
 with $[C_O]=D_+\cap D_-$ is homogeneous,
respectively. It is also true for the entire Beltrami-\dS/\AdS\
space globally. %chart by chart.
 And the generator sets
$\{T^{\frak{d}_\pm}\}=(\H^\pm, \P_i^\pm, \K_i, \J_i)$ of
(\ref{eq:SO14,23}), i.e.%
\be\nno%
& H^\pm :=\partial_t\mp \nu^{2}t x^\nu\partial_\nu,\ \,\, {\mathbf
P}^\pm_i :=\partial_i\mp l^{-2}x_i
 x^\nu\partial_\nu,&\\\label{eq:dpm}%
 & \K_i:=t \partial_i -c^{-2} x_i
 \partial_t,\ \ \J_i=\frac{1}{2}\epsilon_i^{\,jk}L_{jk}, &%
\ee%
 span the \dS/\AdS-doublet $(\frak{d}_+,\frak{d}_-)$ listed in Table I, respectively.

{ It is straightforward to calculate its Christoffel connection
 for the Beltrami-\dS/\AdS\ space.
Namely,
\be\label{ChrsB}%
 \Gamma^{B\lambda}_{\pm
\mu\nu} (x)= \pm l^{-2}\sigma_{\pm}^{-1}(x) ( \delta^\lambda_\mu
x_\nu +
\delta^\lambda_\nu x_\mu),%
 \ee%
which is obviously metric compatible. %
And it is straightforward  to get the Riemann, %curvature,
Ricci
 %curvature
  and scale curvature, respectively, as follows
 \be\nno
&R^{B\mu}_{\pm \nu \la \si}(x)=\pm
l^{-2}(g^B_{\pm\nu\la}\dl^\mu_\si-g^B_{\pm\nu\si}\dl^\mu_\la)&\\\label{riccicurB}
&R^B_{\pm \mu \nu }(x)%= R^\la_{\pm \mu \nu \la}
= \pm 3 l^{-2}g^B_{\pm\mu\nu}, \,~~\,R^B_\pm(x) = \pm  12 l^{-2}.&%
\ee %
 Then it is  the positive/negative constant curvature
Einstein manifold with $\Lambda_\pm=\pm 3 l^{-2}$,
respectively.

In addition, the generators in the sets
$\{T^{\frak{d}_\pm}\}=(\H^\pm, \P_i^\pm, \K_i, \J_i)$ of the
\dS/\AdS\ algebra $\frak{d}_\pm$ can be regarded as Killing
vectors of Beltrami-\dS/\AdS\ metric (\ref{eq: BKmetric}). And
with respect to  these Killing vectors the  Lie derivatives vanish
for the metric, connection and curvature.

Further, it is straightforward to check that the geodesic motion
of metric (\ref{eq: BKmetric}) is indeed the inertial motion
(\ref{eq:uvm}) of Newton's law of inertia as was shown in
\cite{Lu,LZG,BdS}.

It is interesting to see that   if origin lightcone equation %$C_O$
(\ref{eq:CO}) is shifted   by $\mp l^2$, Eqs %$\frak{D}_\pm$
(\ref{eq:domains}) can be reached  as a pair of related regions on
\Mink-space $M$ with boundaries  (\ref{eq: sigma=0}) as a
`pseudosphere',
 respectively. This has been done long ago by Minkowski and others
  (see, e.g. \cite{Synge}). If the flatness of
 space is relaxed,  they are just the domain conditions and absolutes  for the
Beltrami-\dS/\AdS\ spaces $D_\pm$ of \dS/\AdS-invariance. They
indeed  share the same origin lightcone structure $[C_O]$ from
(\ref{eq:Blightcone}) and (\ref{eq: BKmetric}), i.e.
$[C_O]=D_+\cap D_-$
 and  ${\cal L}={\cal D}_+\cap {\cal D}_-$ at common origin. So, the \dS/\AdS\ \SR\
 on $D_\pm$ form a double
 $[{\cal D}_+,
{\cal D}_-]_{D_\pm}$.

\subsection{Special Relativity Triple and Relativistic
Quadruple}

It is important that both Beltrami-\dS/\AdS\ spaces $D_\pm$ share
the $[C_O]$ with \Mink-space $M$ as the tangent space at origin to
both $D_\pm$ so that together with the Poincar\'e double,
\dS/\AdS\ double, , and \dS/\AdS\ doubles, there are also the
\dS/\AdS-${\cal P}$ doubles and the \dS/\AdS-${\cal P}_2$ doubles
$[{\cal D}_\pm, {\cal P}_2]_{D_\pm/M_\pm}$ with Einstein manifold
$M_\pm$, respectively. Then, within $IM(1,3)\sim PGL(5,R)$ there
are naturally  six doubles and they form the
 quadruple ${\frak
Q}_{PoR}$  with common Lorentz isotropy  and $R_\pm=M\cap
M_\pm\cap D_+\cap D_-$. But, as far as  %
cosmological constants $\Lambda_\pm$ are concerned, only two
\dS/\AdS-${\cal P}_2$ doubles $[{\cal D}_\pm, {\cal
P}_2]_{D_\pm/M_\pm}$ are consistent kinematically in principle.

It should be emphasized that these different geometries are
automatically combined together within  \LFT s (\ref{eq:LFT}) of
the inertial motion group $IM(1,3)\sim PGL(5,R)$ based on the
\PoRcl. And the homogeneous Lorentz group ${\cal L}$ of common
isotropy is just a
  subgroup of $IM(1,3)$ for all four relativistic kinematics.

{It should also be noticed that these combined structures can also
be reached even by contraction approach\cite{IW} under limit of
$l\to 0$
or $\Lambda\to \infty$ with respect to a suitable contraction procedure
 based on the \PoRcl. %However,
In fact, for the cases of kinematics the contraction %approach
based on the \PoRcl\  more reasonable approach is to introduce a
dimensionless contraction parameter $\epsilon$ and to replace $l$ by
$\epsilon l$, then taking the limit of $\epsilon\to 0$. It can be
shown that under this contraction, the \dS/\AdS-groups and their
invariant Beltrami-\dS/\AdS\ spaces $D_\pm$ with common Lorentz
group isotropy  just contract to the ${\cal P}_2$ group and its
invariant degenerate manifold $M_\pm$ for $R_\pm$ (\ref{eq:Rpm}),
which is the contraction form of domain conditions
(\ref{eq:domains}), respectively. While, under the contraction
$\epsilon \to \infty$, both \dS/\AdS-invariant
Beltrami-\dS/\AdS-spaces $D_\pm$ become the ${\cal P}$-invariant
\Mink-space $M$. Thus, even for the contraction approach\cite{IW},
there are still the \SR\ triple on three non-degenerate spacetime
$M/D_\pm$ and a relativistic quadruple in the sense of ${\cal
L}={\cal P}\cap {\cal P}_2\cap {\cal D}_+\cap {\cal D}_-$ and $R_\pm
=M\cap M_\pm\cap D_+\cap D_-$\footnote{ In fact, the so-called
infinite cosmological constant contraction and the dual  Poincar\'e
group had been studied in \cite{l=0}, but it didn't based on the
\PoRcl, and relevant geometric results %are all vanishing, which
are not so meaningful. By means of the contraction, degenerate
metrics (\ref{gpm}) has also been found first in \cite{Huang09}. }.

In fact, this is also make sense physically for %experiments and observations of
the inertial observers, {whose world lines should always be
time-like straightlines. No matter when and where they would
change the inertial frames among four types of inertial
coordinates with respect to four types of time and space
coordinate translations in
(\ref{eq:combinatory}) of common Lorentz isotropy, they may %always
change them. This is very important for their experiments and
observations. Namely, among all relativistic kinematics with
relevant inertial frames to be made up with respect to all
possible time and space transformations and the required Lorentz
isotropy, they should find that although Einstein's \SR\ is
perfect so far for the free space without gravity in ordinary
scale,} at cosmic scale our universe must kinematically prefer the
\dS\ \SR\ with the $[{\cal D}_+,{\cal P}_2]_{D_+/M_+}$ of
$\Lambda_+=3l^{-2}$ and their Robertson-Walker-like counterparts.
%%%%%%%%%%%%%%%%%%%%%%%%%%%%%%%%%%%%%%%%%%%%%%%%%%%%%%%%%%%%%%%%%%%%%%%%
%%%%%%%%%%%%%%
%%%%%%%%%%%%%%%%%%^%%%%%%%%%%%%%%%%%%%%%%%%%%%%%%%%%%%%%%%%%%%%%%%%%%%%
\section{Concluding Remarks}\label{sec: remarks}

It should be emphasized that all kinematics should be based on the
\PoRcl\ and its symmetry. Then the inertial motion group follows
for Newton's law of inertia, i.e. $IM(1,3)\sim PGL(5,R)$.  In
addition to three kinds of Poincar\'e/\dS/\AdS-invariant \SR, with
inertial motion the
 ${\cal P}_2$ kinematics can also be set up based on the
\PoRcl\ on a degenerate Einstein manifold
 $M_\pm$ with  $\Lambda_\pm=\pm 3l^{-2}$
induced from   $C_O$ (\ref{eq:CO}) as absolute for its
space/time-like region $R_\pm$ (\ref{eq:Rpm}), respectively. With
the Lorentz isotropy and the common $R_\pm$, there is the
Poincar\'e double  $[{\cal P},{\cal
  P}_2]_{M/M_\pm}$ and dual to it there is also the \dS\ double $[{\cal D}_+,{\cal D}_-]_{D_\pm}$.
  Together with other  four
 doubles  $[{\cal D}_\pm,{\cal P}]_{D_\pm/M}$ and
 $[{\cal D}_\pm,{\cal P}_2]_{D_\pm/M_\pm}$, there is
 the relativistic quadruple ${\frak Q}_{PoR}$ for three kinds of
\SR\cite{Lu,LZG,1950s,BdS,Duality}  as the \SR\ triple on the
intersected non-degenerate \Mink/\dS/\AdS-space\cite{SRT},
respectively. {All these issues automatically appear in
$IM(1,3)\sim PGL(5,R)$.}

For the fate of our universe, it is possibly  a
Robertson-Walker-\dS\ with $\Lambda_+$. So, the \dS\ \SR\ and associated %\dS-${\cal P}_2$
double $[{\cal D}_+,{\cal P}_2]_{D_+/M_+}$ should be payed much
attention. Since  the ${\cal P}_2$-invariant $M_+$ is just for the
space-like region $R_+$ of the Beltrami-\dS\ space, it may not be
so important at least at classical level.
Whatever, the Robertson-Walker counterpart of the %relativistic
quadruple ${\frak Q}_{PoR}$ can be given. The \dS\ \SR\ with
$[{\cal D}_+,{\cal P}_2]_{D_+/M_+}$ and its Robertson-Walker
version  provide  kinematics at the cosmic scale.

 As was well-known, dynamics %and gravity
should coincide with kinematics. It seems reasonable to expect
that these combinational structures may be shed light on dynamics
of three kinds of special relativity and that of the relativistic
quadruple.

It is worthy to mention that since the symmetry of the very special
relativity (VSR)\cite{vsr} is subgroup of ${\cal P}$, taking into
account of four types of translations in (\ref{eq:H}) and
(\ref{eq:P}) there should be the VSR of ${\cal P}_2$ and the
quadruple ${\frak Q}_{VSR}$ for the VSR triple.

{ It should also be mentioned that for the maximal set of symmetries
in 4d spacetime, like the \Mink\ spacetime, there is a 15-parameter
conformal group $SO(2,4)$. This is also true for the 4d \dS/\AdS\
spacetime, respectively (see, e.g. \cite{C3}). However, as was just
mentioned this group contains always the special conformal
transformations that are not \LFT s and cannot preserve the inertial
motion (\ref{eq:uvm}).  It is still relevant to the group
$IM(1,3)\sim PGL(5,R)$ in the sense that there are 10-parameter
kinematic subgroups in both them, although the whole conformal group
is not a subset of the group $IM(1,3)$. Of course, one may consider
the conformal extension of $IM(1,3)\sim PGL(5,R)$ that should
contain the conformal extension of the relativistic quadruple
${\frak Q}_{PoR}$.

It should be noticed first as far as some basic concepts and all
principles are concerned, the approach to kinematics based on the
\PoRcl\ and its symmetry, i.e. the \LFT s of $IM(1,3)\sim PGL(5,R)$,
are completely different from that of Einstein's \GR. The relation
of our approach with \GR\ is a very important issue. We will explore
it elsewhere.

 It should be noticed first as far as some basic
concepts and all principles are concerned, the approach to
kinematics based on the \PoRcl\ and its symmetry, i.e. the \LFT s of
$IM(1,3)\sim PGL(5,R)$, are completely different from that of
Einstein's \GR. For example, there is no invariant metric for the
whole \LFT s of $IM(1,3)\sim PGL(5,R)$, rather there are metric for
kinematics of ten generators including either that of Lorentz
isotropy or of space isotropy. While in \GR, the pseudo-Riemann
metric for spacetime is assumed to exist from beginning. In fact,
from the beginning Einstein denied the \PoR\ and started from his
equivalence principle and general covariance principle
\cite{1922}. Although the \Mink\ space $M$, % is a flat Einstein's manifold,
the Beltrami-\dS/\AdS\ space $D_\pm$ of radius $l$, % is an Einstein manifold with $\Lambda_\pm=\pm3 l^{-2}$,
and  the %dual  Poincar\'e group
${\cal P}_2$-invariant degenerate manifold $M_\pm$ all % of $R_\pm$
 are
 Einstein manifolds of constant curvature,
respectively, the cosmological constants correspond to each of
them are different, except for the
 pairs of $(D_+, M_+)$ and $(D_-, M_-)$. In addition, in view of
\GR, as long as the spacetime is curved, there should be no (global)
inertial motions (\ref{eq:uvm}) so that it is hard to explain why
there are (global) inertial motions (\ref{eq:uvm})
in the  $M_\pm$ and the %Beltrami-\dS/\AdS\ spacetime
$D_\pm$ of constant curvature, respectively. On the other hand,
although \LFT s (\ref{eq:LFT}) may be viewed as a particular type of
(differentiable) arbitrary coordinate transformations  in view of
Einstein's general covariance principle, it is hard to explain in
\GR\ why these \LFT s form such a
 group $IM(1,3)\sim PGL(5,R)$ that keeps inertial motions invariant, which completely departs
 from Einstein's original intention.}
In fact, there is no gravity for  the group $IM(1,3)\sim PGL(5,R)$
based on the \PoRcl\ and for all kinematics. How to introduce
gravity in view of the \PoRcl, of course, is also very important. In
order to describe gravity consistently with symmetry of the \PoR\ in
the localized version, it should be
 based on the local-globalization of the
corresponding \PoR\ with full kinematic symmetry of 1+3d spacetime
(see, e.g. \cite{Duality,PoI}). {Namely, to localize the kinematic
symmetry first, then to connect the localized ones patch by patch
globally with transition functions valued at the full symmetry.}

 We will  study further these issues.
%%%%%%%%%%%%%%%%%%%%%%%%%%%%%%%%%%%%%%%%%%

%%%%%%%%%%%%%%%%%%%%%%%%%%%%%%%%%%%%%%%%%

\begin{acknowledgments}
This work is partly completed in ``Connecting Fundamental Physics
with Observations" programm, KITPC, CAS. We would like to thank Z.
Chang, Y.B. Dai, B.L. Hu, Z.N. Hu, C.-G. Huang, Q.K. Lu, J.P. Ma,
X.A. Ren, X.C. Song, Y. Tian, S.K. Wang, K. Wu, X.N. Wu, Z. Xu,
M.L. Yan, H.X. Yang, C.Z. Zhan, X. Zhang, B. Zhou and C.J. Zhu for
valuable discussions. The work is partly supported by NSFC Grants
No. 10701081, 10975167.

\end{acknowledgments}

\end{document}